\documentclass[11pt]{article}

\usepackage{graphics}

\usepackage{cite}

\def\code#1{{\tt #1}} %
\def\abbrev#1{{\sc #1}} %
\def\ppnum#1{\begin{flushright}\vspace{-13mm}#1\end{flushright}} %
\def\ppdatenum#1#2{\begin{raggedright}#1\\#2\end{raggedright}} %
\def\pptitle#1{\begin{center}\LARGE{#1}\\ \vspace{2cm}\Large Leif
  L{\"o}nnblad\\ \vspace{1cm}\normalsize Theory Division, CERN\\CH-1211
  Gen{\`e}ve 23, Switzerland\\
  E-mail: \code{Leif.Lonnblad@cern.ch}\end{center}} %

\def\abbrev#1{{\sc #1}}

\def\jetset{\abbrev{Jetset}} %
\def\pythia{\abbrev{Pythia}} %
\def\lepto{\abbrev{Lepto}} %
\def\ariadne{\abbrev{Ariadne}} %
\def\h1{\abbrev{H1}} %
\def\cteq2l{\abbrev{CTEQ2L}} %

\def\as{\alpha_S} %
\def\tordas{${\cal O}(\as)$} %
\def\q2{Q^2} %
\def\tq2{$\q2$} %
\def\w2{W^2} %
\def\tw2{$\w2$} %
\def\tx{$x$}
\def\Tsx{Small-$x$}
\def\tsx{small-$x$}

\def\k2t{k_\perp^2}
\def\tk2t{$\k2t$}
\def\m2t{m_\perp^2}
\def\tm2t{$\m2t$}
\def\kt{k_\perp}
\def\tkt{$\kt$}
\def\mt{m_\perp}

\def\p2t{p_\perp^2}
\def\tp2t{$\p2t$}

\def\et{E_\perp}
\def\tet{$\et$}
\def\f2{F_2}
\def\tf2{$\f2$}
\def\ee{e^+e^-}
\def\tee{$\ee$}
\def\qq{q\bar{q}}
\def\tqq{$\qq$}

\def\tqb{$\bar{q}$}
\def\t3{$3$}
\def\tb3{$\bar{3}$}

\def\gev2{\mbox{GeV}^2}
\def\tgev2{$\gev2$}

\def\gaeq{\,\lower3pt\hbox{$\buildrel > \over\sim$}\,}
\def\laeq{\,\lower3pt\hbox{$\buildrel < \over\sim$}\,}
\def\tdy{Drell--Yan}
\def\n2c{N_c^2}
\def\tn2c{$\n2c$}
\def\g2qq{g\rightarrow\qq}
\def\tg2qq{$\g2qq$}
\def\req#1{~(\ref{#1})}
\def\refig#1{~\ref{#1}}
\def\fcap#1{\caption[dummy]{{\it #1}}}

\begin{document}

\begin{titlepage}

  \ppnum{CERN--TH/95--212}
  \vspace{1cm}
  \pptitle{\Tsx\ Effects in $W$ + jets Production at the Tevatron}
  \vspace{2cm}
  \begin{abstract}

    The jet structure in events with \tdy-produced $W$ bosons is
    discussed, and a new model for describing such event is presented.
    The model is shown to explain recent measurements of the $W$--jet
    rapidity correlation and predicts a transverse energy flow
    at high $W$ rapidities (corresponding to probing \tsx\ partons in
    one of the incoming protons) higher than conventional parton cascade
    event generators.

  \end{abstract}

    \vspace{1cm}

  \begin{center}
    {\it Submitted to Nucl.\ Phys.\ B}
  \end{center}

  \vspace{1cm} %

  \ppdatenum{CERN--TH/95--212}{August 1995}

\end{titlepage}

\section{Introduction}
\label{secintro}

The description of \tsx\ partons within hadrons has attracted a great
deal of interest, especially after the measurements of the proton
structure function $F_2$ at \tx\ values down to $10^{-4}$ at HERA,
where a substantial increase was found \cite{H1struct93,ZEUSstruct93}.
Although the \tsx\ rise was first predicted by the so-called BFKL
evolution equation \cite{BFKL77:1,BFKL77:2} it soon turned out
\cite{BallForte941,BallForte942} that it could also be explained in
terms of the conventional Altarelli--Parisi (DGLAP) evolution
equations \cite{DGLAP:1,DGLAP:2,DGLAP:3,DGLAP:4}. Instead, much of the
focus has been directed towards the study of hadronic final states in
deep inelastic lepton--hadron scattering (DIS) events at small \tx,
and it has been suggested that the large flow of transverse energy in
the proton direction found in such events is a signal of BFKL dynamics
\cite{MARTINetflow94}.

Much can also be learned from comparing data with different models
implemented in Monte Carlo event generators. So far it has been shown
that generators built around a conventional DGLAP-inspired
initial-state parton showers, such as \pythia\ \cite{JETPYT94,JETPYT93}
and \lepto\ \cite{LEPTO91}, with strong ordering in virtuality,
completely fail to describe things like the forward transverse energy
flow at small \tx, while a generator such as \ariadne\ \cite{ARIADNENOW}
-- although not implementing BFKL evolution, but sharing with it the
feature that emissions are unordered in transverse momenta --
describes such event features quite well \cite{H1flow94,H1flow95}.

Besides deep inelastic lepton--hadron scattering, \tdy\ production in
hadron--hadron collisions is one of the cleanest probes of hadronic
structure. Recent results from the D0 collaboration \cite{GEOFFcdm94} at
the Tevatron shows a surprising feature of events with \tdy-produced
$W$ bosons, namely the decorrelation in rapidity between the $W$ and
the associated jets. Although the typical \tx-values probed in $W$
events at the Tevatron is on the order of {\small $\sqrt{m_W^2/S}$}
$\approx 80/1800 \approx 0.04$, for large rapidities of the $W$, one
of the incoming partons has a much smaller momentum fraction of the
proton (e.g.\ $y_W \approx 2$ gives $x_1 \approx 6 \times 10^{-3}$ and
$x_2 \approx 0.3$). Therefore it could be worth while to take the
experience gained from studying \tsx\ final states at HERA and try to
apply it to large rapidity $W$-production at the Tevatron.

In this paper the Dipole Cascade Model (DCM)
\cite{CDMinit86,CDMplain88}, on which the \ariadne\ program is built,
is extended to also model the jet structure of \tdy\ production
events. The main feature of the DCM for DIS \cite{CDMdis89} is that
gluon emission is treated as final state radiation from the colour
dipole formed between the struck quark and the proton remnant as in
fig.\refig{figdisdip}a. In this way there is no explicit initial state
radiation, and the proton structure enters only in the way the dipole
radiation is suppressed due to the spatial extension of the proton
remnant. This approach has some problems when it comes to describing
features particular to the initial state, such as the initial splitting
of a gluon into a \tqq-pair.

\begin{figure}
  \setlength{\unitlength}{0.07mm}
  \begin{picture}(2000,600)(-100,0)
    \thicklines
    \put(200,125){\line(1,0){600}}
    \thinlines
    \put(200,150){\line(1,0){200}}
    \put(400,150){\vector(1,2){50}}
    \put(400,150){\line(1,2){100}}
    \put(500,350){\line(2,1){300}}
    \put(500,350){\vector(2,1){150}}
    \qbezier(500,350)(450,350)(450,390)
    \qbezier(450,390)(450,430)(400,430)
    \qbezier(400,430)(350,430)(350,470)
    \qbezier(350,470)(350,510)(300,510)
    \qbezier(300,510)(250,510)(250,550)
    \qbezier(250,550)(250,590)(200,590)
    \put(450,550){\makebox(0,0){$\gamma/Z^0$}}
    \qbezier(850,450)(900,300)(850,150)
    \put(850,450){\vector(-1,4){0}}
    \put(850,150){\vector(-1,-4){0}}
    \thicklines
    \put(1200,125){\line(1,0){600}}
    \put(1200,575){\line(1,0){600}}
    \thinlines
    \put(1200,150){\line(1,0){200}}
    \put(1400,150){\vector(1,2){50}}
    \put(1400,150){\line(1,2){100}}
    \put(1500,350){\line(-1,2){100}}
    \put(1500,350){\vector(-1,2){50}}
    \put(1400,550){\line(-1,0){200}}
    \multiput(1500,350)(50,0){5}{\line(1,0){25}}
    \put(1650,300){\makebox(0,0){$W$}}
    \qbezier(1825,150)(1875,350)(1825,550)
    \put(1825,150){\vector(-1,-4){0}}
    \put(1825,550){\vector(-1,4){0}}
    \put(400,50){\makebox(0,0){(a)}}
    \put(1400,50){\makebox(0,0){(b)}}
  \end{picture}
  \fcap{The colour dipoles that initiate the dipole cascade
    in (a) DIS and (b) \tdy\ production of $W$.}
  \label{figdisdip}
\end{figure}

The simplest extension of the DCM to also treat \tdy\ production would
be to treat gluon emission as final-state radiation from the colour
dipole formed between the two remnants, as in fig.\refig{figdisdip}b.
However, as is seen from fig.\refig{figordalp}, the leading order
$W$+jet diagrams all correspond to initial-state radiation (except for
the last one, which is the least important). And in particular it is
clear that if the gluon emission is treated as final-state radiation
between the two remnants, it would be difficult to explain the
contribution to the transverse momentum of the $W$ from the
diagrams in fig.\refig{figordalp}.

\begin{figure}
  \setlength{\unitlength}{0.07mm}
  \begin{picture}(2000,600)(-100,0)
    \put(0,50){\line(1,1){150}}
    \put(150,200){\line(0,1){200}}
    \put(150,400){\line(-1,1){150}}
    \multiput(150,400)(40,40){4}{\begin{picture}(70,70)
      \qbezier(0,0)(30,30)(60,0)
      \qbezier(60,0)(70,-10)(60,-20)
      \qbezier(60,-20)(50,-30)(40,-20)
      \qbezier(40,-20)(10,10)(40,40)\end{picture}}
    \multiput(150,200)(40,-40){4}{\begin{picture}(20,20)
      \qbezier(0,0)(10,-10)(20,-20)\end{picture}}
    \put(0,550){\vector(1,-1){75}}
    \put(150,400){\vector(0,-1){100}}
    \put(150,200){\vector(-1,-1){75}}
    \put(355,300){\makebox(0,0){{\large +}}}
    \put(355,20){\makebox(0,0){(a)}}
    \put(400,50){\line(1,1){150}}
    \put(550,200){\line(0,1){200}}
    \put(550,400){\line(-1,1){150}}
    \multiput(550,200)(40,-40){4}{\begin{picture}(70,70)
      \qbezier(0,0)(30,-30)(60,0)
      \qbezier(60,0)(70,10)(60,20)
      \qbezier(60,20)(50,30)(40,20)
      \qbezier(40,20)(10,-10)(40,-40)\end{picture}}
    \multiput(550,400)(40,40){4}{\begin{picture}(20,20)
      \qbezier(0,0)(10,10)(20,20)\end{picture}}
    \put(400,550){\vector(1,-1){75}}
    \put(550,400){\vector(0,-1){100}}
    \put(550,200){\vector(-1,-1){75}}
    \multiput(1000,40)(40,40){4}{\begin{picture}(70,70)
      \qbezier(0,0)(30,30)(60,0)
      \qbezier(60,0)(70,-10)(60,-20)
      \qbezier(60,-20)(50,-30)(40,-20)
      \qbezier(40,-20)(10,10)(40,40)\end{picture}}
    \put(1160,200){\line(0,1){200}}
    \multiput(1160,400)(40,40){4}{\begin{picture}(20,20)
      \qbezier(0,0)(10,10)(20,20)\end{picture}}
    \put(1160,400){\line(-1,1){160}}
    \put(1320,40){\line(-1,1){160}}
    \put(1000,560){\vector(1,-1){80}}
    \put(1160,400){\vector(0,-1){100}}
    \put(1160,200){\vector(1,-1){80}}
    \put(1370,300){\makebox(0,0){{\large +}}}
    \put(1370,20){\makebox(0,0){(b)}}
    \multiput(1420,140)(40,40){4}{\begin{picture}(70,70)
      \qbezier(0,0)(30,30)(60,0)
      \qbezier(60,0)(70,-10)(60,-20)
      \qbezier(60,-20)(50,-30)(40,-20)
      \qbezier(40,-20)(10,10)(40,40)\end{picture}}
    \put(1580,300){\line(-1,1){160}}
    \put(1780,300){\line(-1,0){200}}
    \multiput(1780,310)(40,40){4}{\begin{picture}(20,20)
      \qbezier(0,0)(10,10)(20,20)\end{picture}}
    \put(1940,140){\line(-1,1){160}}
    \put(1420,460){\vector(1,-1){80}}
    \put(1580,300){\vector(1,0){100}}
    \put(1780,300){\vector(1,-1){80}}
  \end{picture}
  \fcap{The leading-order Feynman diagrams contributing to $W$ + jet
    production corresponding to the annihilation (a) and Compton (b)
    diagrams.}
  \label{figordalp}
\end{figure}

In previous work \cite{CDMdispom94} the DCM was extended to also
include the boson--gluon fusion diagram in DIS, which can be viewed as
a special case of initial-state gluon splitting into \tqq. In this
paper, this approach is further developed into a general inclusion of
initial-state gluon splitting into \tqq. Also, a way is presented of
taking into account the contribution to the transverse momentum of the
$W$ from the gluon emission, formulated it terms of radiation from the
colour dipole between the two proton remnants.

The layout of this paper is as follows. In sections \ref{secglu} and
\ref{seccmp} the treatment of gluon emission and initial state gluon
splitting is presented. In section \ref{secres}, results for the
$W$--jet rapidity correlation from the improved DCM is compared to a
leading-order calculation and with the conventional DGLAP-inspired
initial-state parton shower approach of \pythia. Also some predictions
are given for the transverse energy flow in high rapidity $W$ events
at Tevatron energies. Finally, in section \ref{secsum}, the
conclusions are presented.

\section{Gluon Emission}
\label{secglu}

The DCM for \tee\ annihilation and deep inelastic lepton--hadron
scattering is described in detail in refs.\
\cite{CDMinit86,CDMplain88,CDMdis89} and only a brief summary of the
features important for this paper will be given here.

The emission of a gluon $g_1$ from a \tqq\ pair created in an \tee\
annihilation event can be described as radiation from the colour
dipole between the $q$ and \tqb. A subsequent emission of a softer
gluon $g_2$ can be described as radiation from two independent colour
dipoles, one between the $q$ and $g_1$ and one between $g_1$ and
\tqb. Further gluon emissions are given by three independent dipoles
etc.

In DIS, the gluon emission comes from the dipole stretched between the
quark, struck by the electro--weak probe, and the hadron remnant. The
situation is the same as in \tee\ above, except that, while $q$ and
\tqb\ are both point-like in the case of \tee, the hadron remnant in
DIS is an extended object.  In an antenna of size $l$, radiation with
wavelengths $\lambda \ll l$ are strongly suppressed. In the DCM, this
is taken into account by only letting a fraction
\begin{equation}
  a=\mu/k_\perp
  \label{eqremfrac1}
\end{equation}
of the hadron remnant take part in the emission of a gluon with
transverse momentum \tkt, where $\mu$ is a parameter corresponding to
the inverse (transverse) size of the hadron.

The phase space available in dipole emission is conveniently pictured
by the inside of a triangle in the $\kappa$--$y$ plane, where
$\kappa\equiv\ln{\k2t}$ and $y$ is the rapidity of the emitted gluon
as in fig.\refig{figphase1}. In these variables the dipole emission
cross section also takes a particularly simple approximate form:
\begin{equation}
d\sigma\propto\alpha_Sd\kappa dy.
\label{eqMEdip1}
\end{equation}
In DIS, assuming that the hadron is coming in with momentum (using
light-cone coordinates) ($P_+,0,\vec{0})$, and is probed by a virtual
photon $(-Q_+,Q_-,\vec{0})$, the triangular area comes from the
trivial requirement
\begin{eqnarray}
  k_{+g} & \equiv & \kt e^y < P_+\nonumber\\
  k_{-g} & \equiv & \kt e^{-y} < Q_-.
\end{eqnarray}
The condition that only a fraction of the remnant participates in an
emission means that
\begin{equation}
  k_{+g}<(\mu/\kt)P_+
\label{eqphasecut}
\end{equation}
and translates into an extra cutoff in the phase space corresponding
to the thick line in fig.\refig{figphase1}. This should be compared to
the initial-state parton shower scenario, where gluon emission is
given by
\begin{equation}
  d\sigma_{q}=\frac{2\alpha_S}{3\pi}
  \frac{1+z^2}{1-z}\frac{f_q(x/z)}{f_q(x)}\frac{dz}{z}\frac{dQ^2}{Q^2}.
  \label{eqiniqsplit1}
\end{equation}
Identifying the ratio of structure functions in eq.\req{eqiniqsplit1}
(corresponding to the dotted line of equal suppression in
fig.\refig{figphase1}) with the extra cutoff (\ref{eqphasecut}) in the
DCM, the two models are equivalent in the low-\tkt\ limit.

\begin{figure}
  \setlength{\unitlength}{0.07mm}
  \begin{picture}(2000,700)(-100,0)
    \put(1050,650){\makebox(0,0){$\kappa$}}
    \put(1650,100){\makebox(0,0){$y$}}
    \put(600,50){\vector(-1,0){50}}
    \put(625,50){\makebox(0,0)[l]{{\footnotesize direction of struck quark}}}
    \put(400,100){\vector(1,0){1200}}
    \put(1000,100){\vector(0,1){600}}
    \put(500,100){\line(1,1){500}}
    \put(1500,100){\line(-1,1){500}}
    \thicklines
    \put(1520,100){\line(-2,1){800}}
    \qbezier[40](1300,100)(1100,300)(900,500)
  \end{picture}
  \fcap{The phase space available for gluon emission in DIS (thin
    lines) and the extra restriction due to the extendedness of the
    proton remnant (thick line). The dotted line corresponds to a line
    of equal suppression due to the ratio of parton density functions
    entering into a conventional parton shower scenario.}
  \label{figphase1}
\end{figure}

As mentioned in the introduction, the simplest way of extending the
DCM to describe gluon emissions in \tdy\ events is to describe it as
radiation from the colour dipole between the two hadron remnants. One
problem with this approach is what to do with the transverse recoil
from the gluon emission. In \tee, this recoil is shared by the $q$ and
\tqb. In DIS, since only a fraction of the remnant is taking part in
the emission, only that fraction is given a transverse recoil,
resulting in an extra, so-called recoil gluon \cite{CDMdis89}. The
corresponding procedure for \tdy\ would be to introduce two recoil
gluons, one for each remnant. However in that way it is impossible to
reproduce the transverse momentum of the $W$ as given by the \tordas\
matrix element.

The \tordas\ matrix element for $q+\bar{q}\rightarrow W+g$ production
takes the form \cite{Halzen78}
\begin{equation}
  M^{q\bar{q}\rightarrow Wg} \propto
  \frac{\hat{t}^2+\hat{u}^2+2m_W^2\hat{s}}{\hat{t}\hat{u}},
  \label{eqMEWg1}
\end{equation}
where $\hat{s}$, $\hat{t}$ and $\hat{u}$ are the ordinary Mandelstam
variables satisfying $\hat{s}+\hat{t}+\hat{u}=m_W^2$. In order to
reproduce this in a parton shower scenario, where the gluon is emitted
``after'' the $W$ is produced, this has to be convoluted with the
parton density functions, and the lowest-order $W$-production matrix
element, again convoluted with the relevant parton densities, has to
be factored out. This introduces some ambiguities, which are solved by
assuming that the rapidity of the $W$ is the same before and after the
gluon emission, resulting in the following cross section, expressed in
the transverse momentum \tk2t\ and rapidity $y_g$ of the emitted gluon
\begin{eqnarray}
  \frac{d\sigma_g}{dy_gd\k2t} & = & \frac{2\alpha_S}{3\pi}
  \frac{f_q(x_q')}{f_q(x_q)}
  \frac{f_{\bar{q}}(x_{\bar{q}}')}{f_{\bar{q}}(x_{\bar{q}})} \times \nonumber
  \\ & & \frac{(\k2t+\m2t+\kt\mt e^{\Delta y})^2+(\k2t+\m2t+\kt\mt
    e^{-\Delta y})^2}{(\k2t+\kt\mt e^{\Delta y})(\k2t+\kt\mt
    e^{-\Delta y})(\k2t+\m2t+\kt\mt(e^{\Delta y}+e^{-\Delta y}))},
  \label{eqMEWg2}
\end{eqnarray}
where $\Delta y = y_g-y_W$, $\m2t=\k2t+m_W^2$, $y_W$ the rapidity of
the $W$ and $x_i$ and $x_i'$ the energy--momentum fractions carried by
the incoming partons before and after the gluon emission so that
\begin{eqnarray}
  x_q = \frac{m_W e^{y_W}}{\sqrt{S}}, & & x_q' = \frac{\mt e^{y_W} +
    \kt e^{y_g}}{\sqrt{S}}, \\ x_{\bar{q}} = \frac{m_W
    e^{-y_W}}{\sqrt{S}}, & & x_{\bar{q}}' = \frac{\mt e^{-y_W} + \kt
    e^{-y_g}}{\sqrt{S}},
\end{eqnarray}
assuming the $q$ coming in along the positive $z$-axis and a total
invariant mass of $\sqrt{S}$.

In the limit $\k2t\ll m_W^2$, eq.\req{eqMEWg2} reduces to the simple
dipole emission cross section in eq.\req{eqMEdip1}, so it is clear
that the strategy outlined above is a good leading log approximation.
It also turns out that it is fairly simple to correct the first gluon
emission so that, disregarding the ratios of parton densities,
eq.\req{eqMEWg2} is reproduced.

\begin{figure}
  \setlength{\unitlength}{0.07mm}
  \begin{picture}(2000,700)(-100,0)
    \put(1050,650){\makebox(0,0){$\kappa$}}
    \put(1650,100){\makebox(0,0){$y$}}
    \put(400,100){\vector(1,0){1200}}
    \put(1000,100){\vector(0,1){600}}
    \put(500,100){\line(1,1){500}}
    \put(1500,100){\line(-1,1){500}}
    \thicklines
    \put(1520,100){\line(-2,1){800}}
    \put(480,100){\line(2,1){800}}
    \put(1100,50){\vector(0,1){50}}
    \put(1100,25){\makebox(0,0){$y_W$}}
    \put(1400,400){\vector(-1,0){50}}
    \put(1420,400){\makebox(0,0)[l]{$\ln{m_W^2}$}}
    \put(600,350){\vector(1,0){50}}
    \put(575,350){\makebox(0,0)[r]{$\ln{k_{\perp\max}^2}$}}
    \qbezier[31](800,100)(950,250)(1100,400)
    \qbezier[30](820,100)(965,245)(1110,390)
    \qbezier[29](840,100)(980,240)(1120,380)
    \qbezier[28](860,100)(995,235)(1130,370)
    \qbezier[27](880,100)(1010,230)(1140,360)
    \qbezier[26](900,100)(1025,225)(1150,350)
    \qbezier[25](920,100)(1040,220)(1160,340)
    \qbezier[24](940,100)(1055,215)(1170,330)
    \qbezier[23](960,100)(1070,210)(1180,320)
    \qbezier[22](980,100)(1085,205)(1190,310)
    \qbezier[21](1000,100)(1100,200)(1200,300)
    \qbezier[20](1020,100)(1115,195)(1210,290)
    \qbezier[19](1040,100)(1130,190)(1220,280)
    \qbezier[18](1060,100)(1145,185)(1230,270)
    \qbezier[17](1080,100)(1160,180)(1240,260)
    \qbezier[16](1100,100)(1175,175)(1250,250)
    \qbezier[15](1120,100)(1190,170)(1260,240)
    \qbezier[14](1140,100)(1205,165)(1270,230)
    \qbezier[13](1160,100)(1220,160)(1280,220)
    \qbezier[12](1180,100)(1235,155)(1290,210)
    \qbezier[11](1200,100)(1250,150)(1300,200)
    \qbezier[10](1220,100)(1265,145)(1310,190)
    \qbezier[9](1240,100)(1280,140)(1320,180)
    \qbezier[8](1260,100)(1295,135)(1330,170)
    \qbezier[7](1280,100)(1310,130)(1340,160)
    \qbezier[6](1300,100)(1325,125)(1350,150)
    \qbezier[5](1320,100)(1340,120)(1360,140)
    \qbezier[4](1340,100)(1355,115)(1370,130)
    \qbezier[3](1360,100)(1370,110)(1380,120)
    \qbezier[2](1380,100)(1385,105)(1390,110)
  \end{picture}
  \fcap{The phase space available for gluon emission in $W$ production
    (thin lines) and the extra restriction due to the extendedness of
    the proton remnants (thick lines). The shaded triangle corresponds
    to the phase space area covered by the $W$.}
  \label{figphase2}
\end{figure}

The ratio of parton densities in eq.\req{eqMEWg2} is instead
approximated by the suppression of the phase space introduced for DIS
in eq.\req{eqphasecut}, which in this case corresponds to suppressions
on both sides of the triangle, as in fig.\refig{figphase2}. One problem
with this procedure is that the \tk2t\ of the gluon and hence of the
$W$ is limited by this suppression to
\begin{equation}
  \k2t<\mu\sqrt{S/4},
\end{equation}
which, with $\mu\approx 1$ GeV, gives $\kt\laeq 30$ GeV. To be able to
describe high-\tkt\ $W$ production, it is clear that the sharp cutoff
in fig.\refig{figphase2}, which in any case is an oversimplification,
must be replaced by a smooth suppression. In \cite{CDMdis89} it was
shown that introducing a power suppression in the disallowed regions
in fig.\refig{figphase1} does not influence the general event shapes
in DIS; it is clear, however, that such a power suppression would
greatly influence the high-\tkt\ spectrum of the $W$.

First, however, the way of obtaining a transverse momentum of the $W$
in the gluon emissions must be formalized. It is clear that, in the
first emission, the gluon corresponds unambiguously to initial state
radiation, and hence the \tkt\ of the gluon must be balanced by the
\tkt\ of the $W$. In further emissions this is not the case, as the
dipole radiation is a coherent sum of the emission from the incoming
partons and the outgoing, previously radiated, gluon. It is therefore
argued that only gluon radiation that takes place close to the $W$ in
phase space should be able to influence the \tkt\ of the $W$; only
gluons emitted in the shaded region of fig.\refig{figphase2},
corresponding to $P_{g+}<P_{W+}$ and $P_{g-}<P_{W-}$, will have their
transverse recoil absorbed by the $W$. Outside this region the
transverse recoil will be treated as in the DIS case above.

\begin{figure}
% GNUPLOT: LaTeX picture with Postscript
\setlength{\unitlength}{0.1bp}
\begin{picture}(2880,1728)(-500,-100)
\put(2334,1314){\makebox(0,0)[r]{DCM gluons $\beta=2$}}
\put(2334,1414){\makebox(0,0)[r]{DCM gluons $\beta=\infty$}}
\put(2334,1514){\makebox(0,0)[r]{LO $W+g$}}
\put(1648,-49){\makebox(0,0){$k_{\perp W}$ (GeV)}}
\put(220,964){%
%
\makebox(0,0)[b]{\shortstack{$d\sigma/dk_{\perp W}$ (pb/GeV)}}%
%
}
\put(2697,151){\makebox(0,0){80}}
\put(2435,151){\makebox(0,0){70}}
\put(2173,151){\makebox(0,0){60}}
\put(1911,151){\makebox(0,0){50}}
\put(1649,151){\makebox(0,0){40}}
\put(1386,151){\makebox(0,0){30}}
\put(1124,151){\makebox(0,0){20}}
\put(862,151){\makebox(0,0){10}}
\put(600,151){\makebox(0,0){0}}
\put(540,1677){\makebox(0,0)[r]{1000}}
\put(540,1202){\makebox(0,0)[r]{100}}
\put(540,726){\makebox(0,0)[r]{10}}
\put(540,251){\makebox(0,0)[r]{1}}
\end{picture}

  \fcap{The transverse momentum spectrum of the $W$ at the Tevatron.
    The full line is the prediction of the \tordas\ $W+g$ matrix
    element as implemented in \pythia\ using CTEQ2L parton density
    functions. The matrix element calculation was cut off at $k_{\perp
      W}=10$ GeV to avoid divergences. The dashed and dotted lines are
    the predictions of the gluon emission in the DCM with
    $\beta=\infty$ and $\beta=2$, respectively.}
  \label{figptWg}
\end{figure}

Figure \ref{figptWg} shows the \tordas\ $W+g$ matrix element
prediction (as implemented in \pythia) of the \tkt\ spectrum of the
$W$ at the Tevatron, compared to the modified DCM described above.
With a sharp cutoff in the phase space, it is clear that the DCM
cannot describe the high-\tkt\ tail of the spectrum. Instead, a smooth
suppression can be introduced by changing eq.\req{eqremfrac1},
allowing a larger fraction $a'$ of the remnant to take part in the
emission with the probability
\begin{equation}
  P(a') \propto
  \frac{\beta(\frac{a'}{a})^\beta}{a'(1+(\frac{a'}{a})^\beta)^2},
\end{equation}
corresponding to a smoothening of the theta function suppression in
fig.\refig{figphase2}, giving a power-suppressed tail. As seen in
fig.\refig{figptWg}, using $\beta=2$ describes well the high-\tkt\
spectrum obtained from the leading-order calculation using the
CTEQ2L\footnote{The CTEQ2L structure function parametrization is used
  in all analyses in this paper where applicable. None of the
  conclusions in this paper were found to be sensitive to this
  choice.} \cite{CTEQ2} structure function parametrization. In the
following, this value of $\beta$ will be used, unless stated
otherwise.

This concludes the description of the $W$ + gluon jet in the DCM.
However, at small $x$, the gluon density in the proton becomes very
large, and the Compton diagrams in fig.\refig{figordalp}b are
dominating.

\section{The Compton Diagrams}
\label{seccmp}

The matrix element for the Compton diagrams looks like \cite{Halzen78}
\begin{equation}
  M^{gq\rightarrow Wq} \propto
  -\frac{\hat{s}^2+\hat{t}^2+2m_W^2\hat{u}}{\hat{s}\hat{t}}.
  \label{eqMEWq1}
\end{equation}
The $s$-channel diagram is, of course heavily suppressed for small
\tkt, and looking only at the $t$-channel diagram, convoluting with
parton densities and factoring out the zeroth order $W$ production
cross section, as in the gluon emission case above, the normal leading
log initial-state parton shower cross section for the splitting of an
incoming gluon into a \tqq\ pair \cite{TSinips88} is obtained:
\begin{equation}
  d\sigma_{q}=\frac{\alpha_S}{4\pi}
  (z^2+(1-z)^2)\frac{f_g(x/z)}{f_q(x)}\frac{dz}{z}\frac{dQ^2}{Q^2},
  \label{eqinigsplit1}
\end{equation}
where $Q^2=-\hat{t}$ and $z=m_W^2/\hat{s}$.

In the DCM, however, there is no initial state gluon splitting into
\tqq, and, just as in the case of final-state \tg2qq\ splitting
\cite{CDMsplit90}, this process has to be added by hand to the DCM.

The simplest way is to introduce the initial-state \tg2qq\ splitting
in the same way as in \cite{CDMsplit90}, as a process competing with
the DCM gluon emission described above. The competition is as usual
governed by the Sudakov form factor using ordering in \tk2t.
Rewriting eq.\req{eqinigsplit1} in terms of the transverse momentum
\tk2t\ and rapidity $y_q$ of the outgoing quark, the probability of
the {\em first} emission to be an initial-state \tg2qq\ splitting at a
certain \tk2t\ and $y_q$ is given by
\begin{eqnarray}
  \frac{dP_{q}(\k2t,y_q)}{d\k2t dy_q} & = &
  \frac{d\sigma_{q}(\k2t,y_q)}{d\k2t dy_q} \times \nonumber \\ & &
  \exp{-\int_{\k2t}^{k_{\perp\max}^2}dk_\perp^{2\prime}\int dy_q'
    \left(\frac{d\sigma_{q}(k_\perp^{2\prime},y_q')}{dk_\perp^{2\prime}
      dy_q'}+\frac{d\sigma_{g}(k_\perp^{2\prime},y_q')}{dk_\perp^{2\prime}
      dy_q'}\right)},
\end{eqnarray}
where the second factor is the Sudakov form factor, corresponding to
the probability {\em not} to have any emission of gluons {\em or} gluon
splittings above the scale \tk2t.

Technically, the extra process is implemented as follows. If the quark
going into the hard interaction on one side is a sea-quark, the
remnant on that side is allowed to ``radiate'' the corresponding
antiquark according to eq.\req{eqinigsplit1}. After such an emission,
the remnant is split in two parts according to the prescription
described in ref.\ \cite{CDMdis89}, one of which forms a dipole with
the ``emitted'' antiquark while the other retains the dipole colour
connection of the original remnant.

If the first emission is a \tg2qq\ splitting, the full \tordas\ matrix
element is used, and the rapidity of the $W$ is assumed to be the same
before and after the emission, as in the gluon emission case. A
splitting later on in the cascade, the kinematic is fixed by requiring
the non-radiating remnant to be unchanged. In all cases, the
transverse momentum of the struck system will of course balance the
\tkt\ of the emitted antiquark. Note that only one initial state
\tg2qq\ splitting is allowed per remnant. This is a good
approximation, since a second such splitting is heavily suppressed by
the parton density functions.

This procedure can be used not only in the case of $W$ production, but
for all processes with a hadron remnant present. In particular it can
be (and is\footnote{This is the default in \ariadne\ version 4.06 and
  later.}) applied in the DIS case.

\begin{figure}
% GNUPLOT: LaTeX picture with Postscript
\setlength{\unitlength}{0.1bp}
\begin{picture}(2880,1728)(-500,-400)
\put(2334,1314){\makebox(0,0)[r]{Full DCM}}
\put(2334,1414){\makebox(0,0)[r]{Leading order}}
\put(1648,1677){\makebox(0,0){(a)}}
\put(1648,-49){\makebox(0,0){$k_{\perp W}$ (GeV)}}
\put(280,914){%
%
\makebox(0,0)[b]{\shortstack{$d\sigma/dk_{\perp W}$ (pb/GeV)}}%
%
}
\put(2697,151){\makebox(0,0){100}}
\put(2278,151){\makebox(0,0){80}}
\put(1858,151){\makebox(0,0){60}}
\put(1439,151){\makebox(0,0){40}}
\put(1019,151){\makebox(0,0){20}}
\put(600,151){\makebox(0,0){0}}
\put(540,1577){\makebox(0,0)[r]{1000}}
\put(540,1246){\makebox(0,0)[r]{100}}
\put(540,914){\makebox(0,0)[r]{10}}
\put(540,583){\makebox(0,0)[r]{1}}
\put(540,251){\makebox(0,0)[r]{0.1}}
\end{picture}
% GNUPLOT: LaTeX picture with Postscript
\setlength{\unitlength}{0.1bp}
\begin{picture}(2880,1728)(-500,-100)
\put(2334,1214){\makebox(0,0)[r]{Full DCM}}
\put(2334,1314){\makebox(0,0)[r]{Parton shower}}
\put(2334,1414){\makebox(0,0)[r]{LO+PS}}
\put(1648,1677){\makebox(0,0){(b)}}
\put(1648,-49){\makebox(0,0){$k_{\perp W}$ (GeV)}}
\put(280,914){%
%
\makebox(0,0)[b]{\shortstack{$d\sigma/dk_{\perp W}$ (pb/GeV)}}%
%
}
\put(2697,151){\makebox(0,0){100}}
\put(2278,151){\makebox(0,0){80}}
\put(1858,151){\makebox(0,0){60}}
\put(1439,151){\makebox(0,0){40}}
\put(1019,151){\makebox(0,0){20}}
\put(600,151){\makebox(0,0){0}}
\put(540,1577){\makebox(0,0)[r]{1000}}
\put(540,1246){\makebox(0,0)[r]{100}}
\put(540,914){\makebox(0,0)[r]{10}}
\put(540,583){\makebox(0,0)[r]{1}}
\put(540,251){\makebox(0,0)[r]{0.1}}
\end{picture}

\fcap{The transverse momentum spectrum of the $W$ at the Tevatron.  In
  (a) the full line is the prediction of the full \tordas\ $W+$jet
  matrix element as implemented in \pythia. The dotted line is the
  prediction of the full DCM with $\beta=2$ and initial-state \tg2qq\
  splitting. In (b) the full line is as in (a) but with the parton
  shower of \pythia\ added after the first emission, the dashed line
  is \pythia\ using only parton showers and the dotted line is the
  same as in (a).}
  \label{figptW1}
\end{figure}

In fig.\refig{figptW1}a, the $W$ \tkt\ spectrum at the Tevatron is
shown, using the full \tordas\ matrix element (as implemented in
\pythia) and using the modified DCM with the initial state \tg2qq\
splitting as implemented in \ariadne\footnote{All results labelled
  \ariadne\ or DCM are actually generated using the zeroth-order $W$
  production in \pythia, with the CDM added and, where indicated,
  using the string fragmentation implemented in \jetset\
  \cite{JETPYT94}.}.  Clearly the DCM does a good job of reproducing
the high-\tkt\ tail of the distribution. In fig.\refig{figptW1}b the
DCM is compared with the two parton shower approaches of \pythia, one
using only parton showers and one using first-order matrix elements
with parton shower added.  Since the DCM is a leading-log cascade,
except that the first emission is uses the full matrix element, it
smoothly interpolates between the pure parton shower description,
which should be a good approximation for small \tkt, and the matrix
element description, which is good for high \tkt, but has to be cut
off at small \tkt\ to avoid divergences.

\section{Results and Predictions}
\label{secres}

In ref.\ \cite{GEOFFcdm94}, it was found that, when looking at the
rapidity of the balancing jet in high-\tkt\ $W$ events at the
Tevatron, no correlation with the $W$ rapidity was found, while a
leading order and a next to leading order calculation predicted a
strong correlation. It was also found that a preliminary
implementation of the DCM model described here reproduced data fairly
well and only gave a very weak correlation\footnote{The results
  presented here differ from the ones in ref.\ \cite{GEOFFcdm94} due
  to a bug introduced in the initial-state \tg2qq\ splitting in the
  preliminary version of \ariadne\ used in that paper.}.

\begin{figure}
                                % GNUPLOT: LaTeX picture with Postscript
  \setlength{\unitlength}{0.1bp}
\begin{picture}(2880,1728)(-500,0)
\put(1607,1334){\makebox(0,0)[r]{DCM}}
\put(1607,1434){\makebox(0,0)[r]{LO+PS}}
\put(1607,1534){\makebox(0,0)[r]{Leading order}}
\put(1648,51){\makebox(0,0){$y_W$}}
\put(280,964){%
%
\makebox(0,0)[b]{\shortstack{$\langle\eta_{\mbox{jet}}\rangle$}}%
%
}
\put(2697,151){\makebox(0,0){2.5}}
\put(2278,151){\makebox(0,0){2}}
\put(1858,151){\makebox(0,0){1.5}}
\put(1439,151){\makebox(0,0){1}}
\put(1019,151){\makebox(0,0){0.5}}
\put(600,151){\makebox(0,0){0}}
\put(540,1677){\makebox(0,0)[r]{1}}
\put(540,1392){\makebox(0,0)[r]{0.8}}
\put(540,1107){\makebox(0,0)[r]{0.6}}
\put(540,821){\makebox(0,0)[r]{0.4}}
\put(540,536){\makebox(0,0)[r]{0.2}}
\put(540,251){\makebox(0,0)[r]{0}}
\end{picture}

\fcap{The average jet pseudo-rapidity $\eta$ vs.\ the rapidity of the
  $W$ $y_W$ at the Tevatron. The jets are reconstructed with a cone
  algorithm using a radius of $0.7$, and in each event a jet with
  $\et>20$ GeV, and $|\eta|<3$ on the oposite side in azimuth
  w.r.t.\ the $W$, is selected. In case of several such jets, the one
  with \tet\ closest to the \tkt\ of the $W$ is chosen.  The full line
  is the leading order calculation as implemented in \pythia, the
  dashed line is the same, but with parton showers and fragmentation
  added, and the dotted line is the full DCM also with fragmentation
  added. (The ``kinkiness'' of the lines are due to limited statistics
  in the simulations.)}
  \label{figWjet1}
\end{figure}

Figure \ref{figWjet1} is an attempt to reconstruct the measurement in
ref.\ \cite{GEOFFcdm94}\footnote{The details in the jet reconstruction
  may differ from that of ref.\ \cite{GEOFFcdm94}. In addition, the
  experimental ambiguity in the $y_W$ determination is not taken into
  account here.} on the generator level. As expected from eqs.\
(\ref{eqMEWg1}) and (\ref{eqMEWq1}), which are both symmetric around
the $W$ rapidity, the leading-order calculation gives a more or less
linear correlation between the mean jet pseudo-rapidity and the
rapidity of the $W$, although $\langle\eta_{\mbox{\tiny
    jet}}\rangle\neq y_q$ due to the smearing of the structure
function convolution and the limited kinematical acceptance for jets.

When parton showers and fragmentation are added to the leading-order
calculation, the smearing is increased. Also, since the phase space
available for emissions is larger on the side where the $x$ of the
incoming parton is smaller, the jets for large $y_W$ are ``dragged''
somewhat towards the centre, destroying the correlation. In the
DCM, this dragging is more pronounced due to the ordering in the
cascade as follows.

\begin{figure}
  \setlength{\unitlength}{0.063mm}
  \begin{picture}(2400,700)(0,0)
    \put(50,100){\vector(1,0){1100}}
    \put(600,100){\vector(0,1){550}}
    \put(600,50){\makebox(0,0){(a)}}
    \put(100,100){\line(1,1){500}}
    \put(1100,100){\line(-1,1){500}}
    \put(750,350){\makebox(0,0){{\footnotesize $\bullet$}}}
    \put(950,450){\makebox(0,0){{\footnotesize $(y_W,\ln{m_W^2})$}}}
    \put(750,350){\vector(-1,-2){30}}
    \put(720,290){\vector(-1,-1){60}}
    \put(660,230){\vector(-2,-1){100}}
    \put(560,180){\vector(-4,-1){150}}
    \put(735,275){\makebox(0,0){{\footnotesize 1}}}
    \put(675,215){\makebox(0,0){{\footnotesize 2}}}
    \put(575,160){\makebox(0,0){{\footnotesize 3}}}
    \put(425,120){\makebox(0,0){{\footnotesize 4}}}
    \put(650,675){\makebox(0,0){{\footnotesize $\kappa$}}}
    \put(1175,70){\makebox(0,0){{\footnotesize $y$}}}
    \put(1250,100){\vector(1,0){1100}}
    \put(1800,100){\vector(0,1){550}}
    \put(1800,50){\makebox(0,0){(b)}}
    \put(1300,100){\line(1,1){500}}
    \put(2300,100){\line(-1,1){500}}
    \put(1300,100){\line(2,1){667}}
    \put(2300,100){\line(-2,1){667}}
    \put(1850,675){\makebox(0,0){{\footnotesize $\kappa$}}}
    \put(2375,70){\makebox(0,0){{\footnotesize $y$}}}
    \put(1950,350){\makebox(0,0){{\footnotesize $\bullet$}}}
    \put(2150,450){\makebox(0,0){{\footnotesize $(y_W,\ln{m_W^2})$}}}
    \put(1950,350){\vector(-1,-3){60}}
    \put(1890,150){\makebox(0,0){{\footnotesize 3}}}
    \put(1890,170){\vector(-1,1){100}}
    \put(1780,290){\makebox(0,0){{\footnotesize 1}}}
    \put(1790,270){\vector(-1,-1){120}}
    \put(1680,130){\makebox(0,0){{\footnotesize 4}}}
    \put(1670,150){\vector(-2,1){100}}
    \put(1550,200){\makebox(0,0){{\footnotesize 2}}}
  \end{picture}
  \fcap{Example of paths, tracing emissions backwards from the hard
    interaction $(y_W,\ln{m_W^2})$ on the \tsx\ side of high-rapidity
    $W$ events for (a) the initial-state parton shower in \pythia,
    where the emissions are ordered both in $x$ and \tk2t, and (b) the
    DCM, where the emissions, although the cascade is ordered in
    \tk2t, when traced backwards from the hard interaction in this
    way, are ordered in $x$ but not in \tk2t.}
  \label{figphase3}
\end{figure}

In the parton shower in \pythia, each step in the backward evolution
of the initial-state shower is ordered in both $x$ and virtuality
\tq2; thus even if the phase space is larger on the \tsx\ side of the
$W$, the shower quickly runs out of phase space due to the ordering in
\tq2\ (resulting also in an ordering in \tk2t) as in
fig.\refig{figphase3}a. The DCM, although ordered in \tk2t, is not
ordered in $x$, or, if the final state partons are traced backwards in
colour from the hard interaction, ordered in $x$ but {\em not} in
\tk2t as in fig.\refig{figphase3}b. In this respect, the DCM is
similar to the BFKL evolution, and it gives a good description of the
large transverse energy flows in \tsx\ events at HERA, which has been
suggested as a signal for the BFKL evolution \cite{MARTINetflow94}.
Because of this, the DCM can better use the increased phase space on
the \tsx\ side of the $W$ and the ``dragging'' effect is larger than for
conventional parton showers in fig.\refig{figWjet1}, and the result
closer to, if not consistent with, the measurement in ref.\
\cite{GEOFFcdm94}.

The increase in transverse energy flow at small $x$ found at HERA
should also be visible at the Tevatron in high-rapidity $W$ events. In
fig.\refig{figEtflow1} the predictions for the \tet\ flow from the
parton shower model of \pythia\footnote{Since no high-\tkt\ jets are
  required and the bulk of the events are at low $k_{\perp W}$, the
  matrix element plus parton shower approach in fig.\refig{figWjet1}
  is not adequate here.} and the DCM of \ariadne\ are shown for
inclusive $W$ events at the Tevatron for two $W$ rapidity intervals.
The two models are fairly similar at central $W$ rapidities, while for
high $y_W$ the CDM gives more transverse energy, despite the fact that
the hard interaction scale ($m_W^2\approx 6400$ $\mbox{GeV}^2$) is much
larger here than at HERA ($\langle Q^2 \rangle \laeq
100$ $\mbox{GeV}^2$).

\begin{figure}
% GNUPLOT: LaTeX picture with Postscript
\setlength{\unitlength}{0.1bp}
\begin{picture}(2160,1728)(100,0)
\put(1633,549){\makebox(0,0)[r]{DCM}}
\put(1633,649){\makebox(0,0)[r]{PS}}
\put(1288,1677){\makebox(0,0){(a) $0.0<y_W<0.5$}}
\put(1288,51){\makebox(0,0){$\eta$}}
\put(400,914){%
%
\makebox(0,0)[b]{\shortstack{$dE_\perp/d\eta$ (GeV)}}%
%
}
\put(1977,151){\makebox(0,0){3}}
\put(1748,151){\makebox(0,0){2}}
\put(1518,151){\makebox(0,0){1}}
\put(1289,151){\makebox(0,0){0}}
\put(1059,151){\makebox(0,0){-1}}
\put(830,151){\makebox(0,0){-2}}
\put(600,151){\makebox(0,0){-3}}
\put(540,1312){\makebox(0,0)[r]{4}}
\put(540,781){\makebox(0,0)[r]{2}}
\put(540,251){\makebox(0,0)[r]{0}}
\end{picture}
% GNUPLOT: LaTeX picture with Postscript
\setlength{\unitlength}{0.1bp}
\begin{picture}(2160,1728)(100,0)
\put(1633,549){\makebox(0,0)[r]{DCM}}
\put(1633,649){\makebox(0,0)[r]{PS}}
\put(1288,1677){\makebox(0,0){(b) $2.0<y_W<2.5$}}
\put(1288,51){\makebox(0,0){$\eta$}}
\put(400,914){%
%
\makebox(0,0)[b]{\shortstack{$dE_\perp/d\eta$ (GeV)}}%
%
}
\put(1977,151){\makebox(0,0){3}}
\put(1748,151){\makebox(0,0){2}}
\put(1518,151){\makebox(0,0){1}}
\put(1289,151){\makebox(0,0){0}}
\put(1059,151){\makebox(0,0){-1}}
\put(830,151){\makebox(0,0){-2}}
\put(600,151){\makebox(0,0){-3}}
\put(540,1312){\makebox(0,0)[r]{4}}
\put(540,781){\makebox(0,0)[r]{2}}
\put(540,251){\makebox(0,0)[r]{0}}
\end{picture}
\fcap{The transverse energy flow in inclusive $W$ events at the
  Tevatron for (a) $0.0<y_W<0.5$ and (b) $2.0<y_W<2.5$. The full and
  dotted lines are the predictions of the DCM in \ariadne\ and of the
  parton shower in \pythia, respectively.}
  \label{figEtflow1}
\end{figure}

In $pp$ collisions we also have to worry about underlying events. In
fig.\refig{figWjet1}, this does not give large effects since a
large-\tet\ jet is required, but for fig.\refig{figEtflow1} the
underlying event would give an extra contribution to the \tet\ flow.
This extra contribution should however be evenly spread out in $\eta$
and independent of $y_W$, and the differences between the parton
shower and DCM approaches should survive. To take this contribution
into account, the multiple interaction model implemented in \pythia\
\cite{TSmult87} has been used. Note, however, that in the case of the
DCM, only the qualitative features of the contribution are completely
relevant, as the parameters of the multiple interaction model probably
need to be retuned to fit the DCM.

\begin{figure}
% GNUPLOT: LaTeX picture with Postscript
\setlength{\unitlength}{0.1bp}
\begin{picture}(2160,1728)(100,0)
\put(1426,1098){\makebox(0,0)[r]{DCM}}
\put(1426,1198){\makebox(0,0)[r]{PS}}
\put(1288,1677){\makebox(0,0){(a)}}
\put(1288,51){\makebox(0,0){$y_W$}}
\put(400,914){%
%
\makebox(0,0)[b]{\shortstack{$dE_\perp/d\eta$ (GeV)}}%
%
}
\put(1977,151){\makebox(0,0){2.5}}
\put(1702,151){\makebox(0,0){2}}
\put(1426,151){\makebox(0,0){1.5}}
\put(1151,151){\makebox(0,0){1}}
\put(875,151){\makebox(0,0){0.5}}
\put(600,151){\makebox(0,0){0}}
\put(540,1388){\makebox(0,0)[r]{6}}
\put(540,1009){\makebox(0,0)[r]{4}}
\put(540,630){\makebox(0,0)[r]{2}}
\put(540,251){\makebox(0,0)[r]{0}}
\end{picture}
% GNUPLOT: LaTeX picture with Postscript
\setlength{\unitlength}{0.1bp}
\begin{picture}(2160,1728)(100,0)
\put(1426,530){\makebox(0,0)[r]{DCM+MI}}
\put(1426,630){\makebox(0,0)[r]{PS+MI}}
\put(1288,1677){\makebox(0,0){(b)}}
\put(1288,51){\makebox(0,0){$y_W$}}
\put(400,914){%
%
\makebox(0,0)[b]{\shortstack{$dE_\perp/d\eta$ (GeV)}}%
%
}
\put(1977,151){\makebox(0,0){2.5}}
\put(1702,151){\makebox(0,0){2}}
\put(1426,151){\makebox(0,0){1.5}}
\put(1151,151){\makebox(0,0){1}}
\put(875,151){\makebox(0,0){0.5}}
\put(600,151){\makebox(0,0){0}}
\put(540,1388){\makebox(0,0)[r]{6}}
\put(540,1009){\makebox(0,0)[r]{4}}
\put(540,630){\makebox(0,0)[r]{2}}
\put(540,251){\makebox(0,0)[r]{0}}
\end{picture}
\fcap{The transverse energy flow two units of rapidity ``behind'' the
  $W$ for inclusive $W$ events, at the Tevatron, i.e.\ for the
  interval $1.0<y_W<1.5$ the \tet\ flow in the pseudo-rapidity
  interval $-1.5<\eta<-1.0$.  The full line is the \pythia\ parton
  shower and the dotted line is the DCM without (a) and with (b)
  multiple interactions.}
  \label{figEtflow2}
\end{figure}

The differences between the parton shower and the DCM approaches are
most significant when the \tet\ flow is measured as a function of
$y_W$ as in fig.\refig{figEtflow2}, where the \tet\ flow, two units of
rapidity away from the $W$, is shown for both models, with and without
the multiple interaction model for the underlying event implemented in
\pythia\ \cite{TSmult87}. It is clear that the underlying event
introduces an extra \tet\ for both models, but that the dependence of
the \tet\ flow on $y_W$ is still different for the two models. As
expected the \tet\ flow is more or less constant for the parton shower
approach, but increases slightly for the DCM due to the increase in
phase space at large $y_W$.

\begin{figure}
% GNUPLOT: LaTeX picture with Postscript
\setlength{\unitlength}{0.1bp}
\begin{picture}(2160,1728)(100,0)
\put(1426,1098){\makebox(0,0)[r]{DCM}}
\put(1426,1198){\makebox(0,0)[r]{PS}}
\put(1288,1677){\makebox(0,0){(a)}}
\put(1288,51){\makebox(0,0){$y_W$}}
\put(400,914){%
%
\makebox(0,0)[b]{\shortstack{$dE_\perp/d\eta$ (GeV)}}%
%
}
\put(1977,151){\makebox(0,0){2.5}}
\put(1702,151){\makebox(0,0){2}}
\put(1426,151){\makebox(0,0){1.5}}
\put(1151,151){\makebox(0,0){1}}
\put(875,151){\makebox(0,0){0.5}}
\put(600,151){\makebox(0,0){0}}
\put(540,1388){\makebox(0,0)[r]{6}}
\put(540,1009){\makebox(0,0)[r]{4}}
\put(540,630){\makebox(0,0)[r]{2}}
\put(540,251){\makebox(0,0)[r]{0}}
\end{picture}
% GNUPLOT: LaTeX picture with Postscript
\setlength{\unitlength}{0.1bp}
\begin{picture}(2160,1728)(100,0)
\put(1426,425){\makebox(0,0)[r]{DCM+MI $\beta=0$}}
\put(1426,525){\makebox(0,0)[r]{ME+PS+MI}}
\put(1426,625){\makebox(0,0)[r]{DCM+MI}}
\put(1426,725){\makebox(0,0)[r]{PS+MI}}
\put(1288,1677){\makebox(0,0){(b)}}
\put(1288,51){\makebox(0,0){$y_W$}}
\put(400,914){%
%
\makebox(0,0)[b]{\shortstack{$dE_\perp/d\eta$ (GeV)}}%
%
}
\put(1977,151){\makebox(0,0){2.5}}
\put(1702,151){\makebox(0,0){2}}
\put(1426,151){\makebox(0,0){1.5}}
\put(1151,151){\makebox(0,0){1}}
\put(875,151){\makebox(0,0){0.5}}
\put(600,151){\makebox(0,0){0}}
\put(540,1388){\makebox(0,0)[r]{6}}
\put(540,1009){\makebox(0,0)[r]{4}}
\put(540,630){\makebox(0,0)[r]{2}}
\put(540,251){\makebox(0,0)[r]{0}}
\end{picture}
\fcap{The transverse energy flow in the interval $-2.5<\eta<-2.0$ for
  $W$ events at the Tevatron as a function of $y_W$. The full line is
  the \pythia\ parton shower and the dotted line is the DCM without
  (a) and with (b) multiple interactions. In (b) the dashed line is
  \pythia\ using \tordas\ matrix elements with addition of parton
  showers and multiple interactions, and the dash-dotted line is the
  DCM with $\beta=\infty$.}
  \label{figEtflow3}
\end{figure}

Similarly, looking at the \tet\ flow in a fixed rapidity interval for
varying $y_W$, the DCM predicts a fairly constant value, while in the
parton shower approach, the flow decreases with increasing $y_W$ as in
fig.\refig{figEtflow3}. To check that the differences are not due to
the fact that the DCM has the correct \tordas\ matrix element in the
first emission, fig.\refig{figEtflow3}b also contains a line with the
matrix element plus parton shower option in \pythia. It is clear that,
although the \tet\ flow is higher since only events with $k_{\perp
  W}>10$ GeV are included, it has the same $y_W$ dependence as the
plain parton shower approach.

Also in fig.\refig{figEtflow3}, the changing of the $\beta$ parameter
in the DCM is shown to have some effect on the \tet\ flow; however,
the dependence on $y_W$ is still the same.

\section{Conclusions}
\label{secsum}

The model presented in this paper is not perfect. The treatment of
initial-state \tg2qq\ splitting is a bit foreign to the original
Colour Dipole Model, and so is the transfer of transverse recoil to
the $W$ for gluon emissions. But as a leading log approximation it
should be just as good as the conventional parton shower approach, and
it has the advantage of correctly describing also high-\tkt\ $W$
production. In addition it gives the opportunity of studying effects
of unordered parton evolution on the hadronic final state, and,
although there are some uncertainties in the overall \tet\ level due
to multiple interactions and the $\beta$ parameter in the DCM, it
would be very interesting to compare the predictions for the $y_W$
dependence of the \tet\ flow presented in this paper with data from
the Tevatron.

Although only $W$ production at the Tevatron has been discussed in
this paper, the model can of course be applied to any \tdy-like
process in any hadron--hadron collision. And, as pointed out above, the
initial-state \tg2qq\ splitting can be used also in deep inelastic
lepton--hadron scattering.

The model presented here is also the ``last piece'' to complete the
DCM description of QCD cascades for all standard processes in \tee,
$ep$ and $pp$ collisions. This is reflected in the fact that \ariadne\
now is fully interfaced to all hard sub-processes in \pythia. However,
some care must be taken when using the multiple interaction model of
\pythia\ together with the DCM, as mentioned above.

\section*{Acknowledgements}

I would like to thank Bo Andersson and G\"{o}sta Gustafson for
valuable discussions.

%\bibliographystyle{physhort}
%\bibliography{references}

\end{document}